\documentclass{article}

\usepackage[T1]{fontenc}
\usepackage{amsmath}
\usepackage{graphicx}
\usepackage{gensymb}
\usepackage{titlesec}
\usepackage{lipsum}
\usepackage{cite}
\usepackage{tikz}
\newcommand*\circled[1]{\tikz[baseline=(char.base)]{\node[shape=circle,draw,inner sep=1pt](char){#1};}}
\usepackage{enumitem}

 \titleformat{\section}
  {\normalfont\scshape}{\thesection}{1em}{}
    
\begin{document}
\title{Aerodynamics or Quantum Collisions: Drag coefficient revision by Schrodinger's equation}
\author{
  Zhao, Maomao\\
  \textit{University of Toronto}\\
  \textit{maomao.zhao@mail.utoronto.ca}
  \and
  Liu, Yufei
 }

\maketitle

\section*{abstract}
Despite the fact that the calculations of drag coefficient and pressure distribution for airfoils can be completed by using Navier-Stoke's equation with help of experimental parameters and advanced computer programming, a simple theoretical approach to these classical problems is still lacked. In this paper we show Schrodinger equation can in fact be a handy tool to describe the mechanics of fluids using rigid sphere in air as an example and further investigate the wave-like properties of fluids. We also provide computational results for simulations of drag coefficient, as well as a comparison to potential theory results of velocity distribution along the surface of a sphere. The final discussion will be focusing on potential generalization of the formulas to other geometrical objects (e.g. airfoils).

\vspace{4.25mm} 

\section*{Theoretical framework}

People have realized the fluid field cannot be fully characterized by viscous and potential theory. The idea that fluids can perform like waves provides us a new perspective while analyzing the picture of fluid fields.
\vspace{4.25mm} 

To describe the wave-like properties of fluids, we first need to find the correspondence between an important quantity in wave mechanics, the product of wave number and radius of the sphere, $ka$, and the classic Reynolds number, $Re$.
\vspace{4.25mm} 

When the radius is constant, the quantity of $ka$ solely depends on the wave number, and $ka \rightarrow 0$ means the wavelength is very big, or if we consider the uniform stream flow of air as a wave, this suggests the ratio of wave length to radius of the sphere is close to infinity. In wave mechanics, the wave number is directly related to momentum. $ka \rightarrow 0$ simply suggests momentum is low, which, in return to fluid field, indicates the system is high in viscosity, i.e. $Re \rightarrow 0$.
\vspace{4.25mm} 

Similarly the product $ka \rightarrow \infty$, momentum is large, corresponds to high velocity in fluids and in other words $Re \rightarrow \infty$ as a consequence.
\vspace{4.25mm} 

Now we modify $Re$ by the generalized $De~Broglie~relation$:

\begin{equation} 
 \rho \overrightarrow{V_{\infty}}=h^*\overrightarrow{k} 
\end{equation} 

Further we let $k=\frac{2\pi}{\lambda_{\infty}}$ and radius of the sphere $a=L$ with $\hbar^*=\frac{h^*}{2\pi}$. Plug in Equation (1) (and the previous alterations) into the expression for $Re=\frac{\rho  V_{\infty}  L}{\mu}$ we have:
\begin{equation} 
 Re=\frac{\hbar^*L}{\mu}\cdot\frac{2\pi a}{\lambda_{\infty}}
\end{equation} 
Thus this shows 
\begin{equation} 
 Re=\frac{\hbar^*}{\mu}\overrightarrow{ka}
\end{equation} 
Given $Re$ and $ka$ are both dimensionless parameters, we see that $\mu$ agree with the generalized Planck constant $\hbar^*$ in base unit. We would discuss later that $\mu$ and $\hbar^*$ agree in order of magnitude as well.

\vspace{4.25mm} 

Now, if we put aside our prejudice and think about the fluid medium as a wave for one second, a rigid sphere in the medium then can serve as an angular symmetric infinite potential well. The fluid gets passed the potential object and endures the collision process by the sphere, or in other words, the absence of a potential would allow the fluid to maintain itself as a "free wave", i.e. a \textit{plane wave} state.

\vspace{4.25mm} 

Here we suggest that the dynamics will be carried by Schrodinger's Equation as in wave mechanics, as we believe the Hamiltonian will capture most of the important characteristics of the flow field for the rigid sphere example. The question whether or not there are other possible operators which can describe the mechanics in much greater precision is beyond the scope of this paper. 

\vspace{4.25mm} 
We use the time-independent nonrelativistic version:

\begin{equation}
\left[\frac{-(\hbar^{\ast})^2}{2 \rho}\bigtriangledown^2+P(x,y,z)\right]\varphi=E_{\infty}\varphi
\end{equation}
\vspace{4.25mm} 

The following analysis and mathematical quantities we seek for will be similar or identical to those in Quantum Scattering Theory, but the fundamental assumptions are not entirely the same. The scattering theory depicts the diagram in terms of collisions of particles, whereas we only care for the explanations by waves. These discrepancies should not cause any contradiction or confusion to readers under the wave-particle duality framework.

\vspace{4.25mm} 

The following is an example of the vector field and streamline simulation for a particular $ka$ value. The rigid ball has constant radius $1$, and the diagrams depict its flow field within the range of radius $r=3$ (to the center of the ball.)

\vspace{4.25mm} 

\textit{Remark.} By the nature of our probabilistic explanation of the wave function, the streamline shall only be regarded as a display. Its physical meaning is not to be confused with a real streamline for classic velocity distributions.

\vspace{4.25mm} 
\begin{figure}[h]
  \setlength{\abovecaptionskip}{0.cm}
  \setlength{\belowcaptionskip}{0.cm}
  \includegraphics[width=\columnwidth]{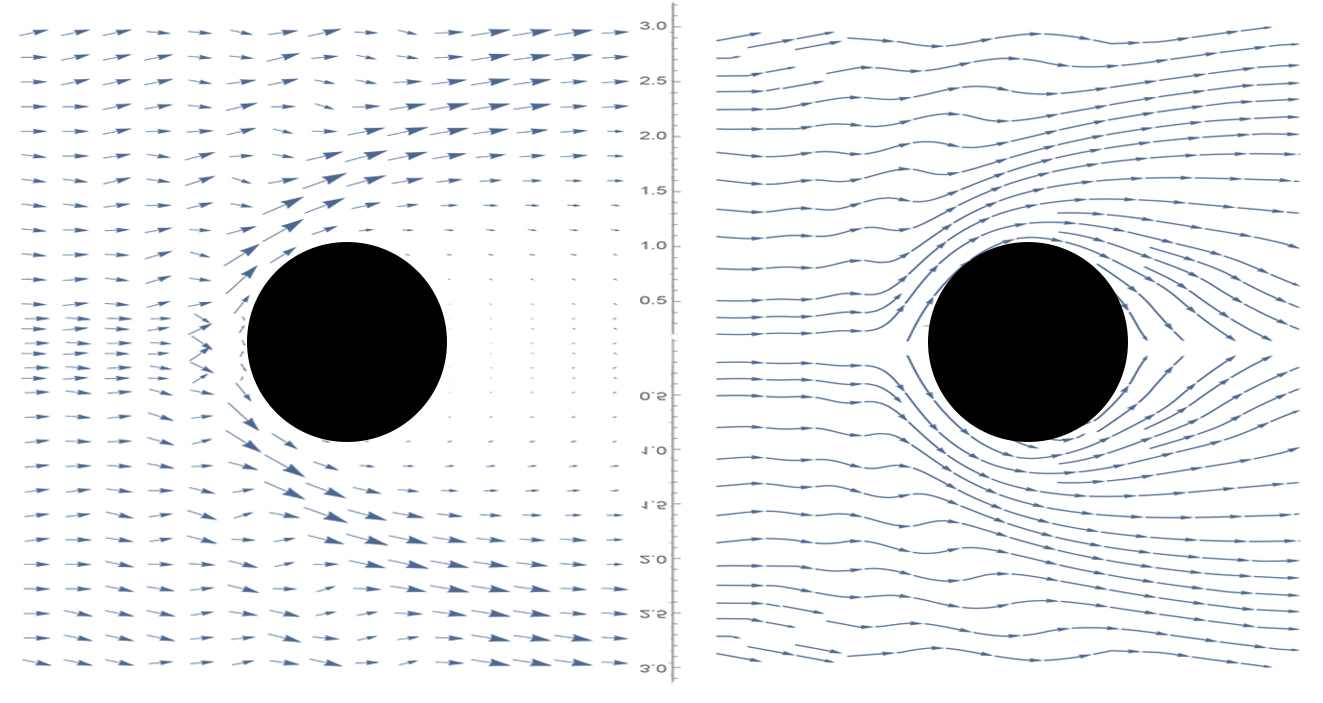}
  \caption{\textit{A simulation for the vector field (left) and streamline diagram (right) of $ka=5$}}
    \label{fig:fig1}
\end{figure}

\vspace{4.25mm} 

\section*{Formulas and Boundary conditions}

We begin with the full Hamiltonian Schrodinger equation for an elastic scattering process: let $<\Phi|$ be the energy eigenket of the kinetic energy operator $H_0$, the Schrodinger's equation is $$(H_0 + V) = E |\Psi>$$ where $V$ stands for the potential (or the scatterer). And we look for the solution $|\Psi >$ which approaches to the free particle solution $|\Phi >$ with the same energy eigenvalue when the scatterer becomes absent.

\vspace{4.25mm} 
Here we may use Lippmann-Schwinger equation instead to avoid the singularity for the operator $\frac{1}{E-H_0}$, then the solution becomes $$| \Psi > = | \Phi > + \frac{1}{E - H_0 + i \epsilon} V |\Psi>$$. In position basis this gives $$< \mathbf{x}| \Psi > = < \mathbf{x}| \Phi> - \frac{2m}{{h^*}^2} \int{d^3 x' \frac{e^{i k |\mathbf{x} - \mathbf{x'}|}}{4 \pi |\mathbf{x}-\mathbf{x'}|} <\mathbf{x'}|V|\Psi>}$$ i.e. the resulting wave function is essentially the original plane wave plus a term representing the effect of the scatterer. The vector $\mathbf{x}$ here is pointed towards the observer's position or where the wave function is evaluated.

\vspace{4.25mm} 
In particular for our rigid ball case, the central potential becomes infinite with finite spherical range, the question simplifies into a separable terms, each with a spatial or an angular factor only. We then employ the partial waves method and the boundary conditions are as follows:

\begin{equation*}
   \begin{cases}
      \Psi \rightarrow e^{i ( \mathbf{p}\cdot \mathbf{r} - E t )/h^*}, & \text{when}\ r\rightarrow \infty\\
      \Psi = 0, & \text{when}\ r \leq a
   \end{cases}
\end{equation*} i.e. we require the wave function degenerate to a plane wave in the far region away from the central potential, and we require the ball to be impenetrable. 

\vspace{4.25mm} 

Thus we obtain the following solution $$< \mathbf{x}|\Psi>=(2\pi)^{-3/2}\sum_{l=0}{i^l(2l+1)e^{i \delta_l} \left[cos~\delta_l~j_l (kr) - sin~\delta_l~n_l (kr)\right] P_l(cos~\theta)}$$ where $l$ is the quantum angular momentum number, $j_l,~n_l$ are the $l^{th}$ term of the first and second kind spherical Bessel functions, $P_l$ is the $l^{th}$ term of the Legendre polynomial, and $\delta_l = arctan \frac{j_l(ka)}{n_l(ka)}$.

\vspace{4.25mm} 

\section*{The tangential velocity on the windward side: Potential theory revised}

Once we obtain the wave function, we can find the vector field $\overrightarrow{J}$ and further we shall pay special attention to the tangential velocity in the near-wall region of the ball, $J_{\theta}$. The following expression gives the explicit form of the vector field: $$\overrightarrow{J} = \overrightarrow{J}(r, \theta) = \mathbf{Re}~-i \Psi^* (\nabla \cdot \Psi) $$ where $\nabla = \frac{\partial}{\partial r} \overrightarrow{r_0}+\frac{\partial}{r \partial \theta} \overrightarrow{\theta_0}$ in spherical coordinates and thus $J_{\theta} = \mathbf{Re}~-i \Psi^* ( \frac{\partial}{r \partial \theta}\Psi) $

\vspace{4.25mm} 
Two examples for the calculation of $J_{\theta}$'s are shown in $Figure~2$ with $ka=3$ from radius $r=a$ to $r=10a$ where $a$ is the radius of the sphere at angles $150\degree$ (left) and $90\degree$ (right).

\begin{figure}[h]
  \setlength{\abovecaptionskip}{0.cm}
  \setlength{\belowcaptionskip}{0.cm}
  \includegraphics[width=\columnwidth]{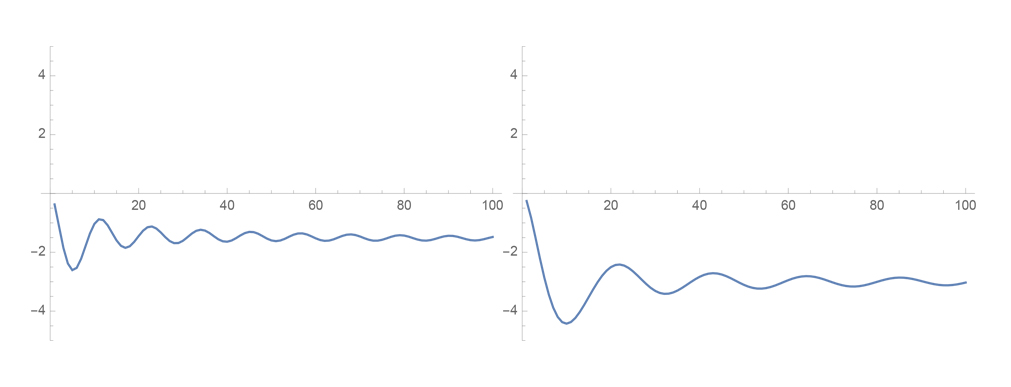}
  \caption{\textit{$J_{\theta}$ for $ka=3$ and radius $a$ to $10a$ at angles $150\degree$ and $90\degree$}}
    \label{fig:fig1}
\end{figure}
\vspace{4.25mm} 

Potential theory has revealed the typical feature of the velocity distribution on the surface of the ball which has contributed into aircraft productions, however, its inaccuracy on the leeward side still remains. Potential theory suggested the maximum velocity on the surface should follow a perfect $sine$ curve(with a scaling of amplitude) hence no pressure difference. Also, the flow field is predicted to be symmetric on both sides of the central sphere. 

\vspace{4.25mm} 
The flow field by wave mechanics shows otherwise. Although agreeing with the windward side distribution which indeed depicts a $sine$ curve tendency, the [leeward] side instead has a "loss in velocity" which must happen in the wake area. (See $Figure~3$)

\begin{figure}[h]
  \setlength{\abovecaptionskip}{0.cm}
  \setlength{\belowcaptionskip}{0.cm}
  \includegraphics[width=\columnwidth]{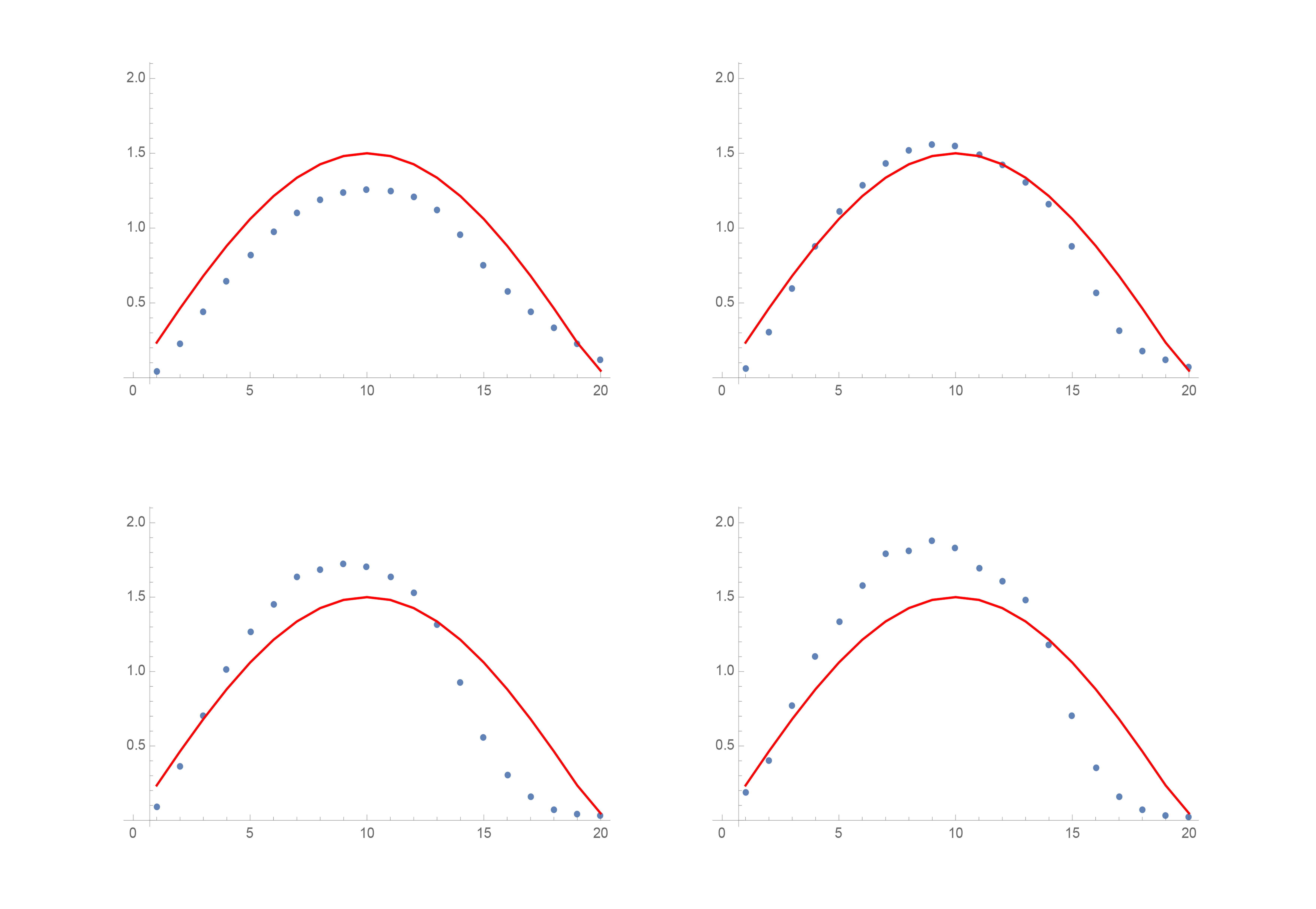}
  \caption{\textit{Comparison of Potential and Wave theory for tangential velocity: $ka=1,~ka=3,~ka=5,~ka=7$}}
    \label{fig:fig1}
\end{figure}

\vspace{4.25mm} 

It matches perfectly with the boundary condition  as well that the velocity right on the surface equals 0 at all angles, whereas Potential theory would give the maximum velocity on surface. Thus the respective radius $r$ which generates the maximum velocity on $J_{\theta}$ under some angle $\theta$ can be considered as a correction to Potential theory too. 

\vspace{4.25mm} 

\section*{Simulation of the Drag coefficient}

For drag force we have the following expression $$F_D=\int_A{\rho u (V_\infty - u)dA}$$, $A$ the area of the control panel, if we let the integral term be a new variable called $V_\infty S^*$ where $V_\infty$ is the far region velocity we would have $$V_\infty S^* = \frac{1}{V_\infty}\int_A{u (V_\infty - u) dA}$$ and thus we obtain $$F_D = \rho V_\infty \cdot V_\infty S^* = \rho V_\infty^2 S^*$$ It's obvious to see that $S^*$ has area in fundamental unit. 

\vspace{4.25mm} 
Now the drag coefficient expression has the form $$C_D = \frac{\texttt{Drag~Force}}{\frac{1}{2}\rho V_\infty^2 S} = \frac{\rho V_\infty^2 S^*}{\frac{1}{2}\rho V_\infty^2 S}$$ i.e. $$C_D = 2 \times \frac{S^*}{S}$$ where $S$ is the projection area of the sphere in the direction of the forwarding flow.

\vspace{4.25mm} 
We shall notice $S^*$ in fact represents the displacement thickness of points in the wake region, thus we would be interested in the calculation of some quantity in wave mechanics which reflects the displacement thickness of such points as in fluid dynamics---the differential cross section in the far region and its near-wall counterpart.
\vspace{4.25mm} 

First notice the differential cross section represents the ratio between the number of scattered particles into a particular solid angle per unit time, and the number of incident particles crossing unit area per unit time, which is an area in base unit. The displacement thickness, in general, is the distance by which the external potential flow is displaced outwards due to the decrease in velocity. Its product with a unit width would also have area as the base unit.
\vspace{4.25mm} 

Since the original differential cross section only describes the statistics for particles in the far away region, here we will use a generalized version of it to investigate the quantity near the surface of the ball. More explicitly, instead of looking at how the incident wave is changed by the spherical outgoing wave far away from the central ball, we develop a similar formula to analyze the difference in incident wave affected by outgoing spherical wave near the wall.

\vspace{4.25mm} 
\begin{equation}
\left| f(\theta)\right|^2 = \left|(\Psi(r,\theta)|_{r \rightarrow \infty} - e^{i k z})/\frac{e^{i k r}}{r} \right|^2
\end{equation}

\vspace{4.25mm} 

\begin{equation}
\left| f'(r,\theta)\right|^2 = \left|(\Psi(r,\theta) - e^{i k z})/\frac{e^{i k r}}{r} \right|^2
\end{equation}

\vspace{4.25mm} 
Comparing equations (5) and (6), one major variation is that the generalized version of differential cross section no longer only depends on the angle factor, it has the spatial factor radius $r$ as a variable too. Although here we will list several examples to show all small $r$ will generate similar curves (even if we take arithmetic average over small $r$'s,) we encourage peer researchers to elaborate more on this end and to further study the stability of this curve over the spatial dependence.

\vspace{4.25mm} 
\begin{figure}[h]
  \setlength{\abovecaptionskip}{0.cm}
  \setlength{\belowcaptionskip}{0.cm}
  \includegraphics[width=\columnwidth]{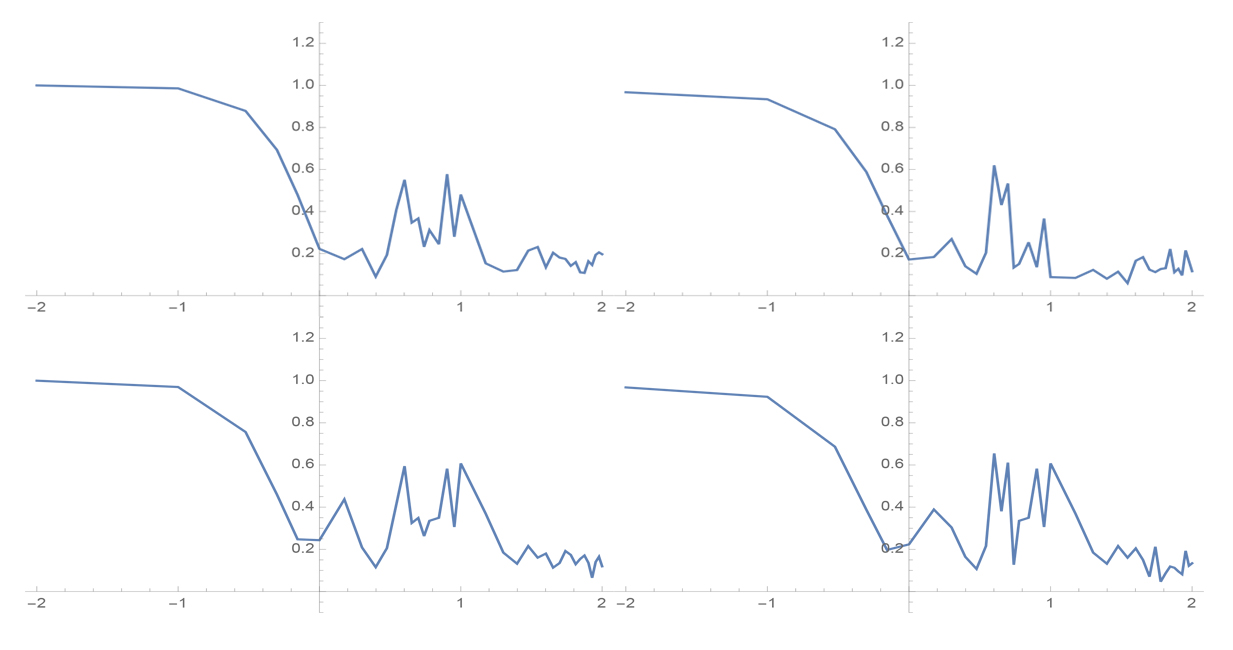}
  \caption{\textit{Drag coefficient simulations without viscosity accounted for $ka=0.01$ to $ka=100$ under $Log10$}}
    \label{fig:fig1}
\end{figure}
These 4 simulations are under different parameter settings: up left one is recorded under $90\degree$ and averaged over radius from $a$ to $2a$; the up right one is under $80\degree$ with the same range of radius; the bottom left one is under $90\degree$ and averaged over $a$ to $3a$; and the bottom right is under $80\degree$ averaged over $a$ to $3a$.

\vspace{4.25mm} 

The constant region for $ka\leq 0.1$ comes from the fact that $low-energy$ partial waves attach to the surface of the ball more evenly, and the entire wave function would be dominated by these low-level partial waves only.

\vspace{4.25mm} 

Now we take account of viscosity into calculation for $C_D$: if the fluid is high in viscosity, the respective oscillation of the wave function becomes negligible. Hence the position where the tangential velocity reaches its maximum grows out to infinity (in response with $Re \rightarrow \infty$) which in return reflects that the thickness of boundary layer grows to infinity. 

\begin{figure}[h]
\begin{center}
  \setlength{\abovecaptionskip}{0.cm}
  \setlength{\belowcaptionskip}{0.cm}
  \includegraphics[width=7cm]{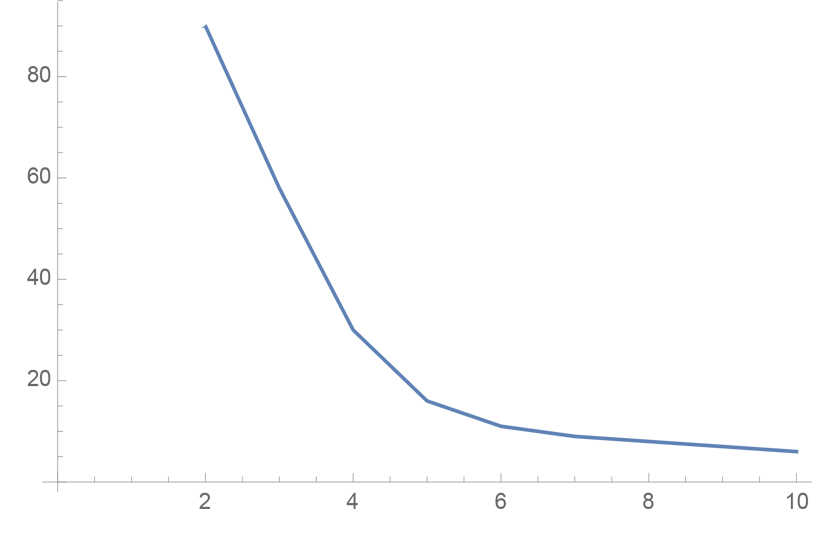}
  \caption{\textit{Boundary layer thickness for $ka=0.01$ to $ka=7$}}
    \label{fig:fig1}
    \end{center}
\end{figure}

\vspace{4.25mm} 

The complete Drag coefficient curve $C_D$ should consist these two types of curves together, i.e. when the viscosity dominates the system ($ka\rightarrow0$), $C_D\propto J_{\theta}$, and when $ka \geq 0.1$, $C_D \propto \left|f'(r,\theta)\right|^2$.

\vspace{4.25mm} 
The following features shown in the diagrams above in fact should help to determine the exact conversion ratio between $Re$ and $\overrightarrow{ka}$:

\begin{enumerate}[label=\protect \circled{\arabic*}]
\item{platform:constant area}
\item{drag crisis}
\item{minimum value}
\item{increasing again with oscillations}
\end{enumerate}

\vspace{4.25mm} 
Here we give one example for the augmented drag coefficient curve by taking $ka=5\times 10^5~Re$ with the value of $\left| f'(r,\theta)\right|^2$ where $\theta=90\degree$ averaged over $r=a$ to $r=3a$. If we denote the boundary layer thickness (or the parameter which directly reflects viscosity) from above by $B$ and we set a pseudo-parameter $\alpha=5\times 10^5$ to enhance the reverse proportional relation between $ka$($Re$) and boundary layer thickness (viscosity), we obtain (very loosely) a formula for the augmented $C_D$ curve:

\begin{equation}
C^*_D=\frac{B}{\alpha \cdot ka}+\left| f'(r,\theta)\right|^2= \frac{B}{5\times 10^5 ka}+\left| f'(r,\theta)\right|^2
\end{equation}
\vspace{4.25mm} 

The following figures will show the comparison between our augmented $C^*_D$ curve and an actual $C_D$ curve from $NASA$ under $Log~10$ basis:

\begin{figure}[h]
\begin{center}
  \setlength{\abovecaptionskip}{0.cm}
  \setlength{\belowcaptionskip}{0.cm}
  \includegraphics[width=7cm]{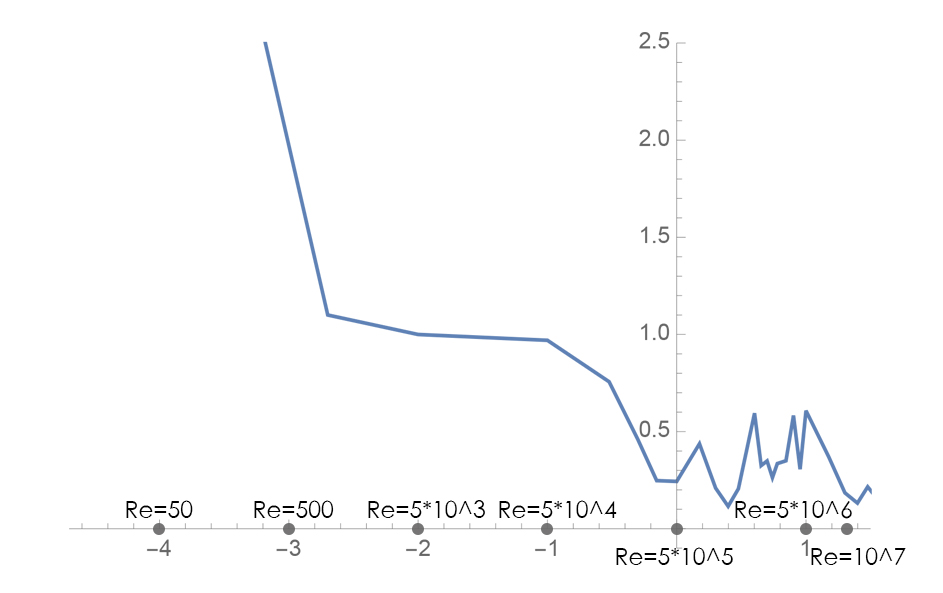}
  \caption{\textit{Augmented $C^*_D$ curve in $Log~10$ base}}
    \label{fig:fig1}
    \end{center}
\end{figure}

\begin{figure}[h]
\begin{center}
  \setlength{\abovecaptionskip}{0.cm}
  \setlength{\belowcaptionskip}{0.cm}
  \includegraphics[width=7cm]{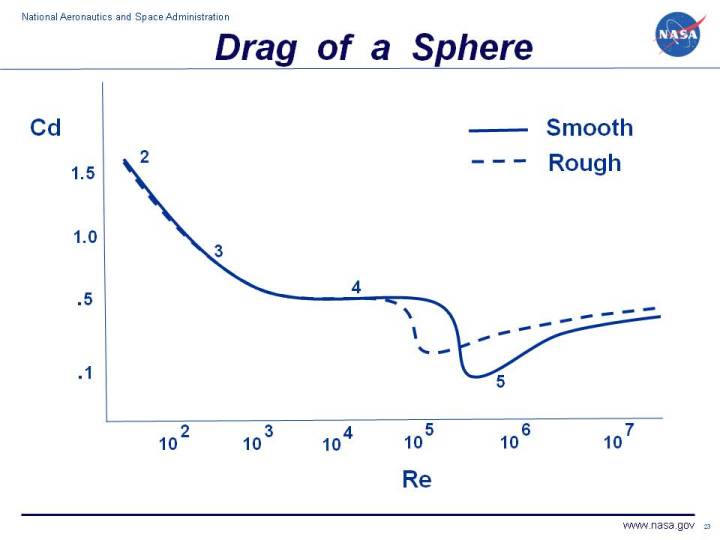}
  \caption{\textit{Drag coefficient curve for rigid ball in air from $NASA$}}
    \label{fig:fig1}
    \end{center}
\end{figure}
\vspace{4.25mm} 

In the Augmented curve diagram, since we've approximated the relationship between $ka$ and $Re$ is: $ka=5\times 10^5~Re$, we only continued the calculation to the value of $ka$'s which are in correspondence with the values of $Re$ from the $NASA$'s diagram. The smallest value for $ka$ in $Log$ base 10 is 0.00002 which should correspond to $Re=10$ and the biggest is $Log_{10}(ka)=1.5$ which corresponds to $Re\approx 1.6\times10^7$.
\vspace{4.25mm} 

The exact value of the Augmented drag coefficient should be scaled by the characteristic value $\frac{\hbar^*}{\mu}$ for air in this particular case, although the specific $\frac{\hbar^*}{\mu}$ values for each fluid medium are still needing to be determined.
\vspace{4.25mm} 

\section*{the Boundary layer evolution and separation}

We adopted a similar approach to study boundary layer evolution when compared with Potential theory, i.e. we analyze the tangential velocity distribution near the surface of the sphere in the flow field at all angles. The following diagram is an example of the velocity distribution for $ka=3$.

\begin{figure}[h]
  \setlength{\abovecaptionskip}{0.cm}
  \setlength{\belowcaptionskip}{0.cm}
  \includegraphics[width=\columnwidth]{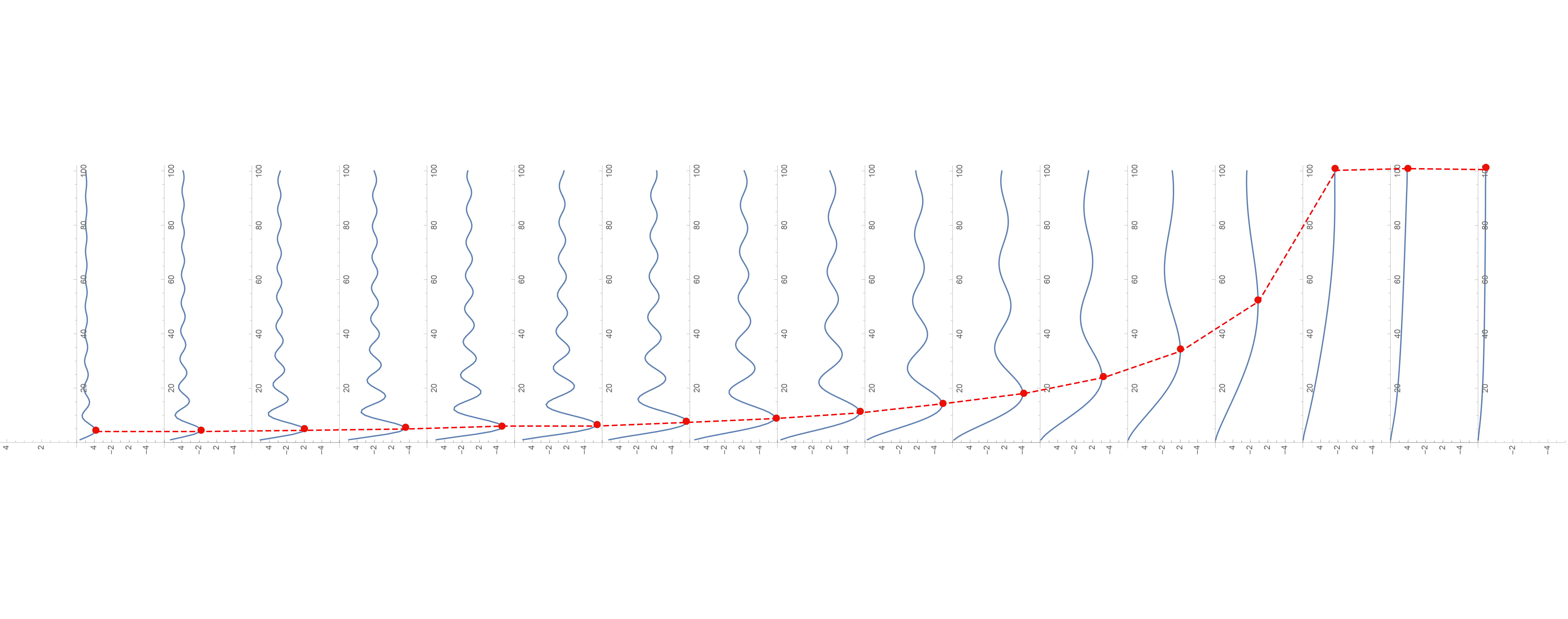}
  \caption{\textit{Tangential velocity distribution at angles between $170\degree$ to $10\degree$ for $ka=3$}}
    \label{fig:fig1}
\end{figure}

\vspace{4.25mm} 
Each velocity curve experiences similar versatile changes: $0$ velocity on surface; smoothly increasing to its maximum value either very close to the surface or far away from the surface (depending on the angles under which the velocity is recorded). In addition, the versatility (the amplitude of the wave) for these curves becomes smaller until the average reaches the theoretical angular component of the wind velocity at boundary (i.e. $r \rightarrow \infty$), i.e. the average value for large radius of these curves decrease to an average amplitude of $V_{\infty}sin~\theta$ (for example for $ka=3$ we have $V_{\infty}sin~90\degree=3$ at $90\degree$.) ($Figure~1$)

\vspace{4.25mm} 
On $Figure~3$ the dotted red line connects all points where the $J_{\theta}$ reach their maximums along the curves. The left hand side of the diagram represents the windward side starting from angle $170\degree$. The red line moves relatively flat on the windward side until around $100\degree$ where the slope of the line increases more rapidly than before. This is a reflection of a growth of both the boundary layer and the displacement thickness.

\vspace{4.25mm} 

The exact thickness for boundary layers of specific fluids, for example air, still needs to be determined later. It relies on the magnitudes of $h^*$, $\rho$, and $\mu$. In other words, information about the characteristic parameters would need detailed discussion since these parameters are directly related to specific fluid mediums.

\vspace{4.25mm} 

\section*{Generalization to airfoil calculations and future studies}

Recall in position basis the full wave function is  $$< \mathbf{x}| \Psi > = < \mathbf{x}| \Phi> - \frac{2m}{h^2} \int{d^3 x' \frac{e^{i k |\mathbf{x} - \mathbf{x'}|}}{4 \pi |\mathbf{x}-\mathbf{x'}|} <\mathbf{x'}|V|\Psi>}$$ and if we replace the rigid ball condition by some other geometrical object that is not spherical symmetric, the original equation no longer can be separated into spatial and angular terms. In other words the differential cross section $f(\mathbf{x},\mathbf{x'})$ and other important parameters would be relying on specific relations between the position vectors $\mathbf{x}$ and $\mathbf{x'}$.

\vspace{4.25mm} 
Let $r=|\mathbf{x}|$, $r'=|\mathbf{x'}|$ and let $\theta$ be the angle between $\mathbf{x}$ and $\mathbf{x'}$. We have $$|\mathbf{x}-\mathbf{x'}| = \sqrt{r^2-2r r' cos~\theta + r'^2 }= r\left(1-\frac{2r'}{r} cos~\theta + \frac{r'^2}{r^2}\right) ^{1/2}$$ and for convenience we denote $g(r,r')=\left(1-\frac{2r'}{r} cos~\theta + \frac{r'^2}{r^2}\right) ^{1/2}$ then $e^{i k |\mathbf{x}-\mathbf{x'}|} = e^{ikr*g(r,r')}$. Similarly $\frac{1}{|\mathbf{x}-\mathbf{x'}|  }= \frac{1}{r*g(r,r')}$.

\vspace{4.25mm} 
Now the solution becomes $$< \mathbf{x}| \Psi > = < \mathbf{x}| \Phi> -\frac{1}{4\pi} \frac{2m}{h^2} \frac{e^{ikr}}{r} \int{d^3 x' \frac{e^{ikr*(g(r,r')-1)}}{g(r,r')} V(\overrightarrow{\mathbf{x'}})<\mathbf{x'}|\Psi>}$$ here we've used the simplification $<\mathbf{x'}|V|\Psi> = V(\mathbf{x'})<\mathbf{x'}|\Psi>$.

\vspace{4.25mm} 

\section*{Final remarks}

As the readers notice the calculation details as well as the diagrams shown in this paper may not suffice to provide highly accurate explanations for the complete fluids over bodies features. This is in fact an entirely new perspective on the macro picture of quantum collision and the paper is attempting to find a more reasonable principle behind aerodynamics other than $NS$ equation. 

\vspace{4.25mm} 

Unfortunately due to the great amount of work we had to do solely for building the framework of this "new method" and limited access to more professional computer calculation/simulation programs, we were unable to elaborate more on parameter settings, etc. Our grids for the vector field/streamline diagrams were merely $20*20$, which may have caused the vortices/turbulence features been evened out. 

\vspace{4.25mm} 

Nevertheless, we hope our dearest peer researchers can identify these fascinating links between fluid mediums and waves and carry on studies in this field with passion.

\vspace{4.25mm} 

\section*{Acknowledgement}

The authors appreciate academic helps from school mates and personal friends Freid Tong, Pratyush Sarkar, Z.F.Zhu,  Department of Mathematics, University of Toronto, and we cherish the love and support from our family.

\vspace{4.25mm} 


\begin{thebibliography}{99}



  \bibitem{Liu} Y.F.Liu  {\em An introduction to Quantum Methods for Flows around a Body}, 2013 (Chinese translation).

  \bibitem{Sakurai} J.J.Sakurai {\em Modern Quantum Mechanics} revised edition, 1994
   
  
  
  \end{thebibliography}
\end{document}